\documentclass[a4paper]{jpconf}
\usepackage{graphicx}
\begin{document}
\title{ On cross-section computation in the 
brane-world models}

\author{Kirpichnikov Dmitry}

\address{Institute for Nuclear Research of the Russian Academy of Sciences,
Prospect of the 60th Anniversary of October 7a, Moscow, Russia, 117312}

\ead{kirpich@ms2.inr.ac.ru}

\begin{abstract}
 We present Mathematica7 numerical simulation of the 
  process $pp\!\rightarrow\!\mbox{jet}\!+\!E_{T}^{miss}$  
  in the 
  framework of modified Randall-Sundrum  brane-world model 
  with one infinite and $n$ compact extra dimensions. 
  We compare the energy missing
  signature with the standard model background
   $pp\rightarrow \mbox{jet}+\nu \bar{\nu}$, which was simulated at
   CompHep. 
  We show that the models with numbers of compact extra 
  dimensions greater than 4 can be probed at the 
 protons center-of-mass energy equal 14 TeV.  We also find
  that testing the brane-world models 
   at 7 TeV on the LHC appears to hopeless.
\end{abstract}

\section{Introduction}
There are a lot of softwares for simulating a 
"New physics" processes at the accelerating experiments.
Such programs as {\bf CompHep} \cite{Boos:2009un}
 and {\bf PYTHIA}~\cite{Sjostrand:2000wi}
are among them. Nevertheless, one can consider 
an infinite extra spatial dimension in the brane-world 
models. {\bf CompHep} and {\bf PYTHIA}  are not adopted for such class
of models. 
In this paper we present the numerical simulations of the 
processes
$pp \rightarrow \mbox{jet}+E_{T}^{miss}$ on {\bf Mathematica7}
in the backgound of modified Randall-Sundrum brane model with
one infinite and $n$ compact extra dimensions 
(RSII-$n$ model). In this brane world model 
neutral particle such as Z boson and photon can leave our
brane, escaping in to the extra dimension of infinite size.
In our sumulation we use Gluck Reya Vogt leading order 
parton distribution functions. In Sec.~\ref{Motiv} we 
 compare LO Gluck Reya Vogt PDF~\cite{Gluck:1998xa}, by
with CTEQ~\cite{Pumplin:2002vw}, 
MRST~\cite{Martin:2002dr} and Alekhin's~\cite{Alekhin:2002fv}
 LO PDFs.
  This PDFs coincide at large QCD
scale parameter squared $Q^2$. In Sec.~\ref{Comparison}
 we compare {\bf CompHep} and {\bf Mathematica7} numerical simulations of the process
$pp \rightarrow \mbox{jet}+Z^0$  
in the framework of standard model.
We discuss RSII-$n$ set up in Sec.~\ref{RSsetup}.
In Sec.~\ref{EnergyMissing} we present numerical 
simulations of the process 
$pp \rightarrow \mbox{jet}+E_{T}^{miss}$
in the framework of RSII-$n$ model.

\section{ Comparison of PDFs at large $Q^2$.
\label{Motiv}}
\begin{figure}[t]
\begin{center}
\includegraphics[width=1\textwidth]{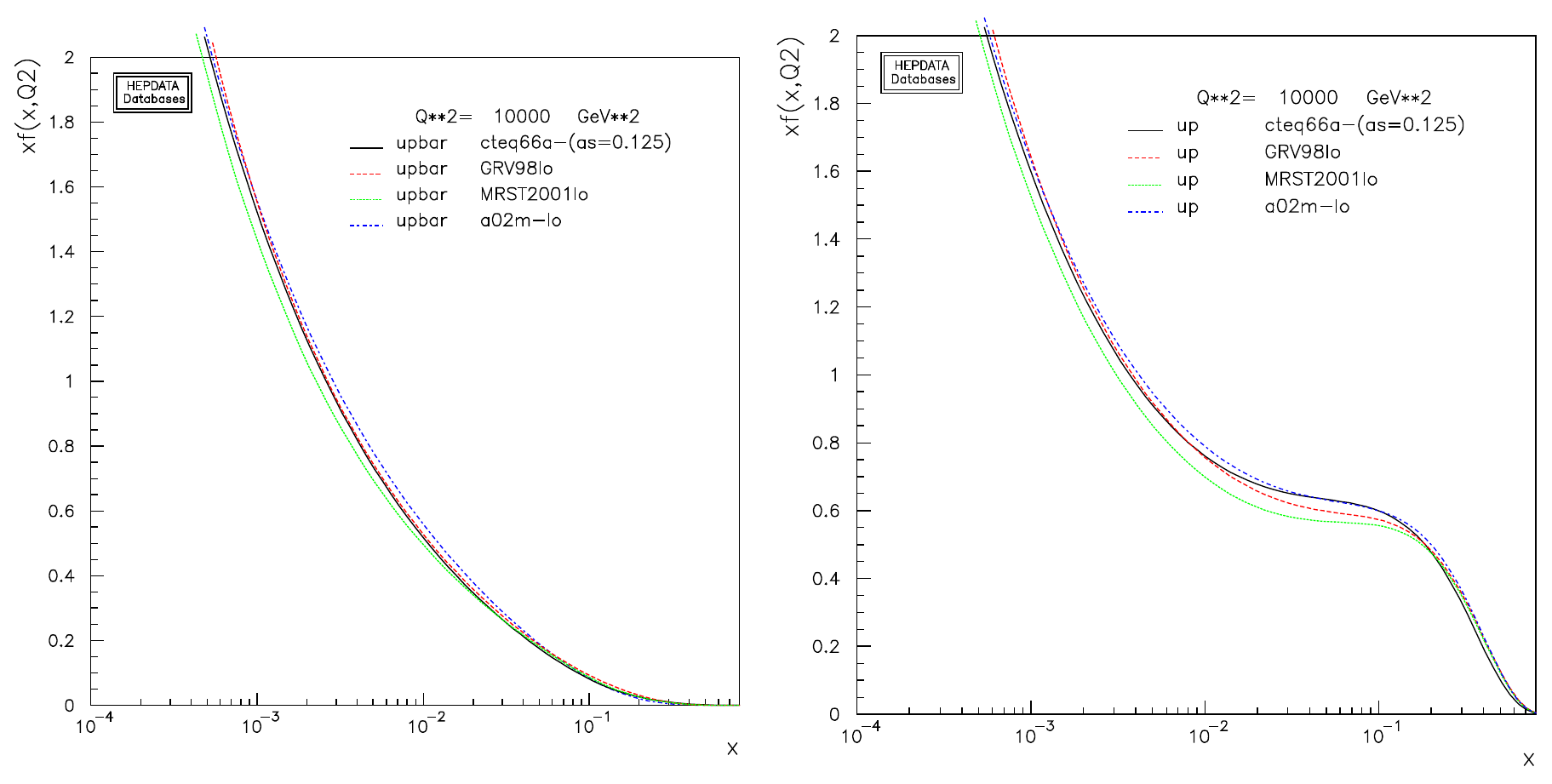}
\caption {LO PDFs $xu(x,Q^2)$ and $x\bar{u}(x,Q^2)$ for
GRV, CTEQ, MRST and Alekhin's collaborations.
 \label{figureA}}
\end{center}
\end{figure}
In this section we compare LO
Gluck Reya Vogt parton distribution functions 
with CTEQ, MRST and Alekhin's PDFs. Corresponding 
distributions for $u$ and $\bar{u}$ quarks are presented 
on Fig.~\ref{figureA}. We suppose that QCD scale parameter is 
fixed at $Q=100$ GeV. These data are taken from 
an open High energy physics database
\cite{hep}.  
We can see from  Fig.~\ref{figureA}
that GRV LO PDFs coincide with CTEQ LO, MRST LO and
Alekhin's LO PDFs at large $Q^2$.
One can 
obtain an analogous distribution for $d$, $\bar{d}$
quarks and gluon. This means that in the numerical analysis 
 we can use GRV LO PDFs 
as well as CTEQ, MRST and Alekhins PDFs if $Q^2$ is fixed at large 
values.   

\section{{\bf CompHep} vs {\bf Mathematica7} in the SM framework.
\label{Comparison}}
\begin{figure}[t]
\begin{center}
\includegraphics[width=1\textwidth]{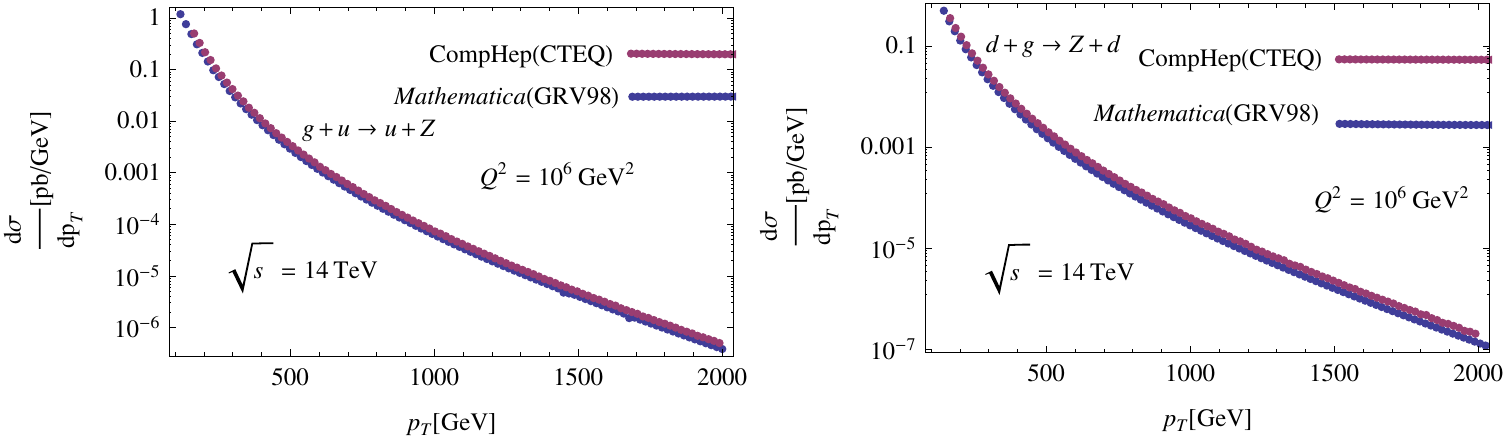}
\caption {Comparison of differential distribution for the
processes $dg\rightarrow Z^0 d$ and $ug\rightarrow Z^0 u$ simualated at {\bf CompHep} and {\bf Mathematica7} in 
the framework of standard model.
 \label{figureB}}
\end{center}
\end{figure}

In this section we 
 compare {\bf CompHep}
and {\bf Mathematica7} numerical simulations for the parton
cross-sections in the framework of standard model. 
In Fig.~\ref{figureB} we show the
 differential
cross-section of the processes
$dg\rightarrow Z^0 d$ and $ug\rightarrow Z^0 u$ versus 
transverse momentum of $Z^0$ boson. The QCD scale is 
fixed at $Q=1000$~GeV. The diagrams for GRV and CTEQ LO PDFs
are coincided. But there is a small discreapancy in 
$x u(x,Q^2)$ distribution at high $p_T$ values. The 
analogous distribution can be obtained for 
the other SM processes such as 
$\bar{d}g\rightarrow Z^0 \bar{d}$, 
$\bar{u}g\rightarrow Z^0 \bar{u}$,
$\bar{d}d\rightarrow Z^0 g$ and
$\bar{u}d\rightarrow Z^0 g$. Nevertheless, the main 
contribution to the process $pp \rightarrow \mbox{jet}+Z^0$
 comes from the gluon cross-sections 
which are shown on Fig.~\ref{figureB}.

\section{Modified Randall-Sundrum model
\label{RSsetup}}
In this section we discuss a peculiar features 
of the modified RSII-$n$ model. Let us consider 
a 3-brane with $n$ compact dimensions embeeded in a $(5+n)$
space-time AdS metric 
\begin{equation}
ds^2=a(z)^2(\eta_{\mu \nu} dx^\mu dx^\nu - \delta_{ij} d \theta^i
d \theta^j) -dz^2 \label{gauge_metric}.
\end{equation}
This metric was suggested by T.~Gherghetta and 
M.~Shaposhnikov \cite{Gherghetta:2000qi}.
 Here $z$ is the infinite extra
dimension, $\theta_{i}$ are the compact extra-dimensions
$\theta_{i}\in[0, 2\pi R_i]$,  $i=\overline{1,n}$, 
$R_i$ are the sizes of compact extra dimensions, 
$n$ - is the number of compact extra dimensions,
$a(z)=e^{-k|z|}$ is a warp factor from Randall-Sundrum model 
and
$k$ is a AdS curvature.

We put entire $SU(2)\times U(1)$ gauge sector, 
as well as the Higgs sector into the bulk space,
but the fermions of the standard modell are 
supposed to be localized on the brane. The action
of the model is
$$
S=\int d^4 x\,dz\prod^n_{i=1} \frac{d\theta_i}{2\pi R_i}\sqrt{g}
\Bigl[-\frac{1}{2}|\,W_{MN}\,|^2+m^2_W
|\,W_M\,|^2-\frac{1}{4}Z^2_{MN}+\frac{1}{2}m^2_Z
Z^2_M
$$
\begin{equation}
-\frac{1}{4}F^2_{MN}+\frac{1}{2}(\partial_M 
\chi)^2-\frac{1}{2} m^2_{\chi}
\chi^2 + \delta(z)\mathcal{L}_{ferm}\Bigr], 
\label{TheAct}
\end{equation}
where $m_W^2=\frac{1}{4}\widetilde{g}_{2}^2 v^2$,
$m_Z^2=\frac{1}{4}
(\widetilde{g}_{2}^2+\widetilde{g}_{1}^2)v^2$
and $m^2_\chi=\lambda v^2$ are the bulk masses of the gauge 
fields 
and Higgs respectively. We also suppose the 
size of the compact extra dimension to be
 $1/R_i \gg \sqrt{s}$. This means that
corresponding KK excitations become infinitely heavy and 
dissappear from the spectrum.
\begin{table}[t]
\begin{center}
\caption{The lower 
bounds on the AdS curvature $k$ for various ~$n$. 
}
\begin{tabular}{ccccccc}
\br
    $n$   & $1$ & $2$ & $3$ & $4$ & $5$ & $6$ \\
    \mr
    \\
    $k , \mbox{GeV} $ & $5.5 \times 10^6$ & $20 \times 10^3$& $2.5 \times 10^3$ &$900$ & $400$  & $300$ \\
\br    
\label{table1}
\end{tabular}
\end{center}
\end{table}

Since the $Z$-boson is not exactly localized on the brane,
one can obtain the bounds on number of compact 
extra dimension $n$ and the AdS curvature $k$. 
We require that the invisible width decay of $Z$-boson
in RSII-n model is bounded by the 
experimental uncertainty of the total 
$Z$-boson width decay~\cite{Amsler}:
$$
\Gamma_{RS}(M_Z)\le \Delta\Gamma^Z_{tot}=1.5 \,\mbox{MeV}.
$$
where
\begin{equation}\Gamma^Z_{RS}=\frac{2\pi}{n\Gamma^2\left(\frac{n}{2}\right)} M_Z
\left(\frac{M_Z}{2k}\right)^n,
\end{equation}
is the invisible decay rate of $Z^0_{bulk}$ boson 
\cite{Kirpichnikov:2012vb}.
These bounds are shown in Tab.~\ref{table1}.
We use these values of $k$ when presenting 
the  numerical simulations in Sec.~\ref{EnergyMissing}

\section{Numerical simulation of the 
process $pp \rightarrow \mbox{jet}+E_T^{miss}$.
\label{EnergyMissing}}
In this section we discuss numerical simulation
of the process $pp \to \mbox{jet}+Z_{bulk}(\gamma_{bulk})$
at the collider experiment in the frame work of 
RSII-$n$ model. Here the jet originates from gluon 
or quark, and $Z_{bulk}$ and $\gamma_{bulk}$ are the 
particles which escape the brane.
\begin{figure}[t]
\begin{center}
\includegraphics[width=0.495\textwidth]
{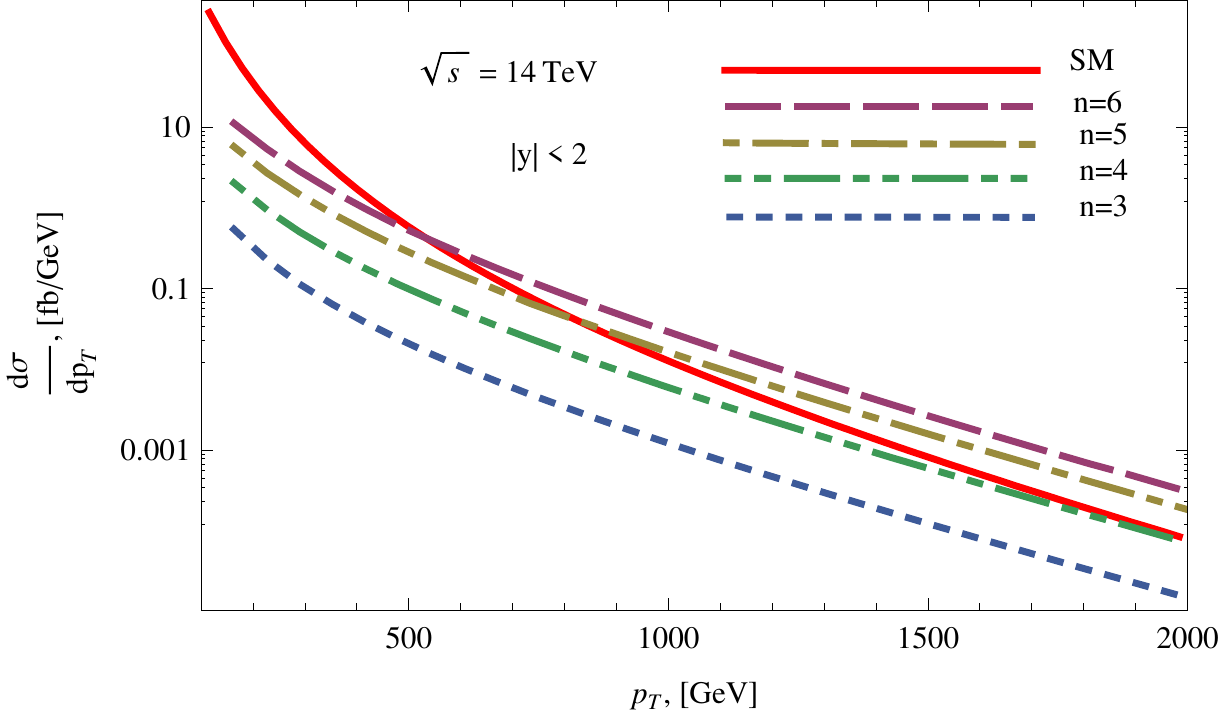}
\includegraphics[width=0.495\textwidth]
{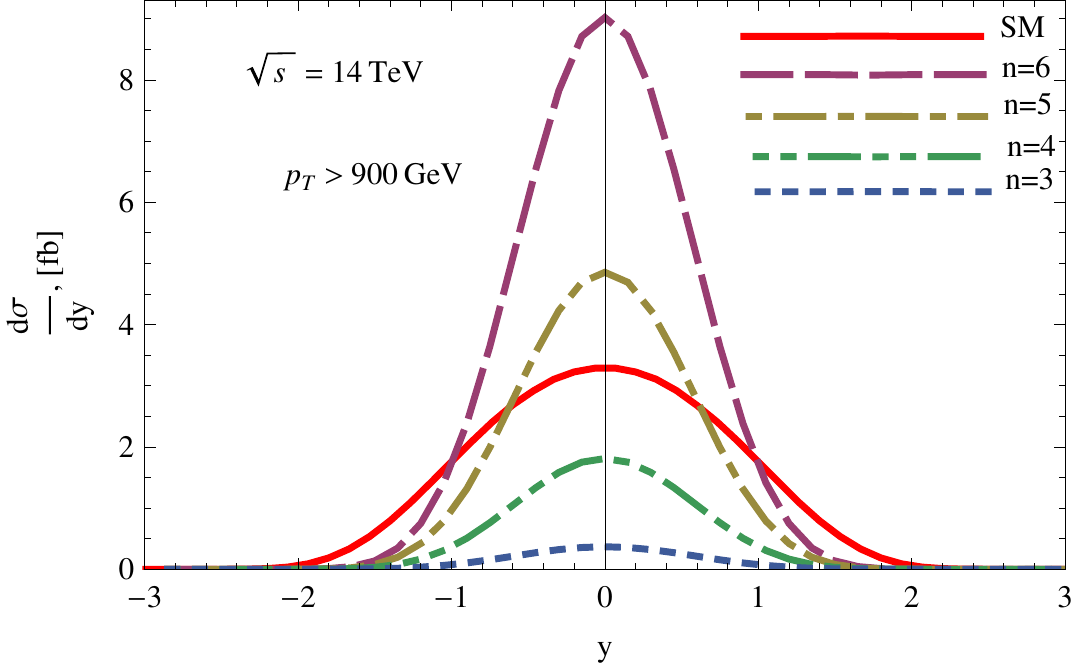}
\caption{  Differential cross section of the process 
$pp \to \mbox{jet}+Z_{bulk}(\gamma_{bulk})$   
for
$n=3,4,5,6$ at the center-of-mass energy of protons equal
 $\sqrt{s}=14$TeV. The
standard model background is  
$pp \to \mbox{jet}+\nu\bar{\nu}$. 
 \label{figure1}}
 \end{center}
\end{figure}  
The differential cross-section of this process is
\begin{equation}
\frac{d^2 \sigma}{dy\, p_T dp_T}\!\!=\!\!\int dx_1 dx_2
\!\!\! \sum_{i,j=q,
\bar{q},g}\!\!\! \frac{f_i(x_1,Q^2)}{x_1}\frac{f_j(x_2,Q^2)}{x_2} 
\frac{1}{4\pi s \,m}\, \overline{\sum_k}|\mathcal{M}_{ij\to k(bulk)}|^2
\label{DiffCrSect}
\end{equation}
where $\mathcal{M}_{ij\rightarrow k (bulk)}$ are 
parton amplitudes for the subprocesses 
$gq\to qZ_{bulk}(\gamma_{bulk})$, 
$\bar{q}q\to gZ_{bulk}(\gamma_{bulk})$ 
(see Ref.~\cite{Kirpichnikov:2012vb} for details);  
$m$ is a bulk invariant mass of 
$Z_{bulk}$ and $\gamma_{bulk}$:
$$
m^2=s x_1 x_2-x_1 p_T \sqrt{s} \e^{-y}-x_2 p_T \sqrt{s} \e^{+y},
$$
 and $f_i(x_1,Q^2)$ are
 GRV LO PDFs \cite{Gluck:1998xa}. The 
QCD scale parameter is fixed at $Q=1$~TeV.
 We compare the distribution (\ref{DiffCrSect})
 with the standard model background 
$pp\to \mbox{jet}+\nu\bar{\nu}$.
This background was computed by {\bf CompHep}
program   \cite{Boos:2009un}.

Let us consider the case of the proton center-of-mass 
energy  equal 14 TeV. In Fig.~\ref{figure1}  we show 
$p_T$ and jet rapidity   
distributions 
of the process $pp\to \mbox{jet}+Z_{bulk}(\gamma_{bulk})$.  
One can see from 
Fig.~\ref{figure1} that if $n=6$ and $n=5$, then the signal  
  dominates over  
 the background for $p_T> 500$ GeV and  $p_T> 750$  GeV,
 correspondingly. 
 And the background dominates over the signal for 
  $n=3,4$.
Once the larger values of $k$ correspond
to the  smaller $n$ (see Tab.~\ref{table1}), 
so that  the  signal cross 
section   grows 
with the increase
of  $n$. It is clear from Fig.~\ref{figure1} that  the jet
 rapidity distributions 
  are correlated with the jet transverse momentum distributions.  
\begin{figure}[t]
\begin{center}
\includegraphics[width=0.495\textwidth]{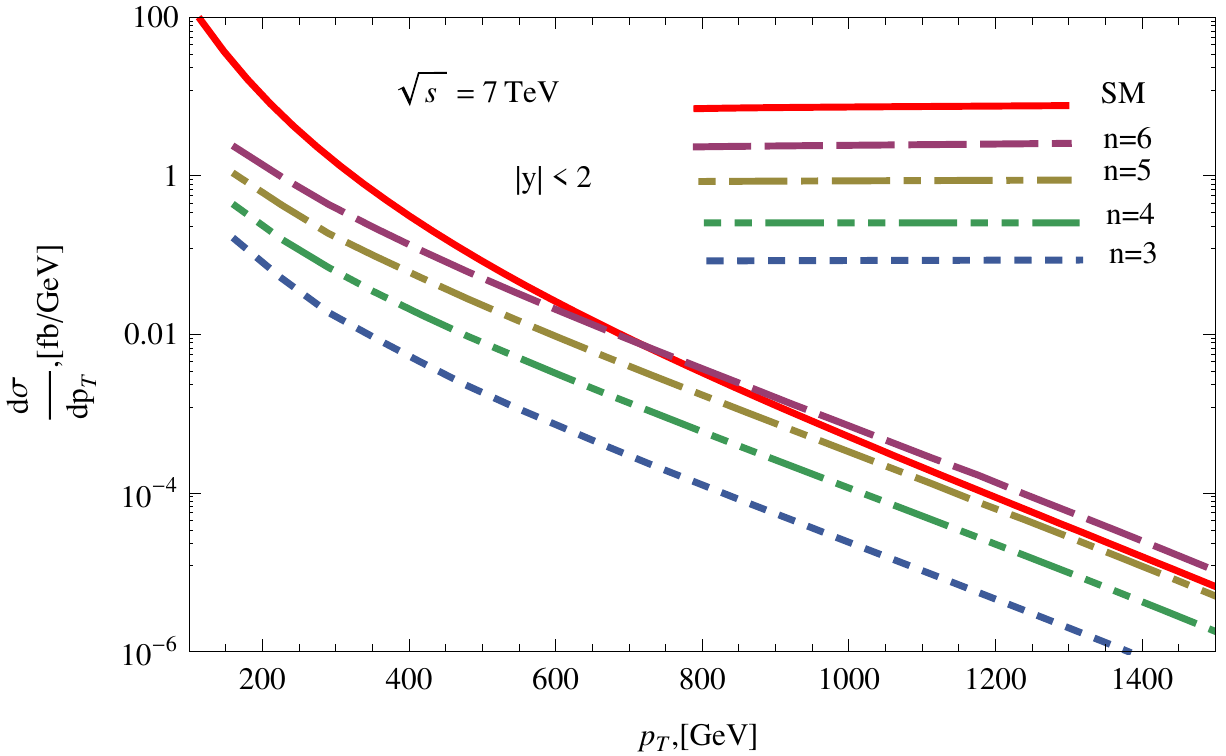}
\includegraphics[width=0.495\textwidth]{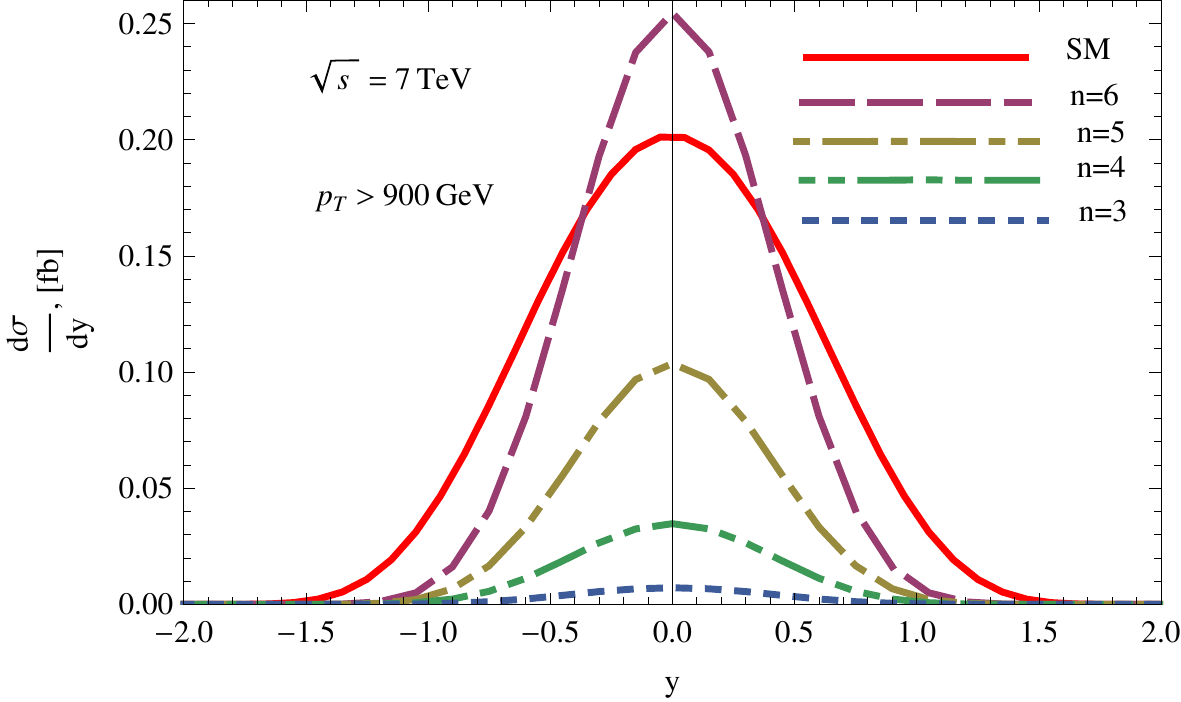}
\caption {  Same as in Fig.~\ref{figure1}, but for
the  energy of protons equal  7~TeV.  
 \label{figure7} }
 \end{center}
\end{figure} 
In Tab.~\ref{table2} we show the integrated 
luminosity and number of signal events needed for
$5\sigma$ discovery at the LHC center-of-mass energy 
equal $14$ TeV. 
Where  the cuts in jet rapidity and transverse momentum
are $|y|<2$,
 $p_T> 900$ GeV. 
 
 \begin{table}
\centering
\caption{
Integrated luminosity $\mathcal{L}$ and numbers of signal events 
$N_S$ for 
various $n$, at the LHC center-of-mass energy $\sqrt{s}=$ 
14 TeV.
}
\begin{tabular}{ccccc}
\br
    $n$   &  $3$ & $4$ & $5$ & $6$ \\
\mr
 $\mathcal{L}, \mbox{fb}^{-1}$ & $7.1 \times 10^2$ & $3.7 \times 10^{1}$& $7.5$ &$3.1$  \\
\mr
 $N_S$ & $ 3.6\times 10^2$ & $10^2$& $ 5\times 10^1$ &$3.8\times 10^1$  \\
\br
    \end{tabular}
\label{table2}
\end{table}
Now let us consider the case $\sqrt{s}=7$ TeV. 
In Fig.~\ref{figure7} 
we show the jet transverse
momentum and rapidity distributions.  Only the case with
$n\ge 6$ can  be probed, since the background
dominates over the signal for $n=3,4,5$.
The 
integrated luminosity required for $5\sigma$  discovery 
at the LHC is~$\mathcal{L}=200\, \mbox{fb}^{-1}$
even for $n=6$.

\section{Summary \label{summary}}
In the framework of RSII-n model the distributions 
 of the process $pp \rightarrow \mbox{jet}+ Z^0_{bulk}(\gamma_{bulk}) $ 
  were simulated 
 on {\bf Mathematica7} with GRV LO PDFs implemented.   
  The detection of the extra spatial dimension at 7 TeV appears to be hopeless. 
  At the LHC energy equal 14 TeV 
 the luminosity needed for $5\sigma$ discovery is in the range
  $(10-100) \mbox{fb}^{-1}$.

\section{Acknowledgements}
We are indebted to A.~B.~Arbuzov and A.~L.~Kataev, 
for helpful  discussions and advices, to ACAT 2013
Organizing Comittee and in particular 
Bin Gong for support and hospitality. 
  This  work  was supported in part by 
grants of  Russian Ministry of  Education 
and  Science NS-5590.2012.2 and
GK-8412, grants of the President of 
Russian Federation MK-2757.2012.2, and grants
of RFBR 12-02-31595 MOL A and RFBR 13-02-01127 A.

\end{document}